\documentclass[aps,prd,twocolumn]{revtex4}
\usepackage{amsmath}
\usepackage{graphicx}
\usepackage{epsfig}

\begin{document}

\title{Can massless QCD dynamically generate heavy quarks?}
\author{Alejandro Cabo-Montes de Oca and Danny Mart\'{\i}nez-Pedrera}

\affiliation{ International Centre for Theoretical Physics, Strada
Costiera 111, Trieste, Italy \\}

\affiliation{ Group of Theoretical Physics, Instituto de
Cibern\'etica, Matem\'atica y F\'{\i}sica, Calle E, No. 309,
Vedado, La Habana, Cuba}

\begin{abstract}
\noindent
 As it was suggested by previous works on a modified
perturbation expansion for QCD, the possibility for the generation
of large quark condensates in the massless version of the theory
is explored. For this purpose, it is firstly presented a way of
well define the Feynman diagrams at any number of loops by just
employing dimensional regularization. After that, the calculated
zero and one loop corrections to the effective potential indicate
a strong instability of the system under the generation of  quark
condensates.  The also evaluated quark condensate dependence of 
particular two loop terms does not modify the instability picture
arising at one loop. The results suggest a possible mechanism for
a sort of  Top Condensate Model to be a dynamically fixed
effective action for massless QCD. The inability of lattice
calculations in detecting this possibility  could be related with
the limitations in treating the fermion determinants.
\end{abstract}

\maketitle

\section{Introduction}

\ The explanation of chiral and flavor properties of QCD is one of
the fundamental research issues in high energy physics
\cite{leut,fritzsch,fritzsch1,lane,colwitt,vafwitt}. A central
problem in this field is the question about the origin of the
quark masses, which indirectly determine  the nature of most of
the observable matter. More generally, the explanation of the
elementary particle mass spectrum is considered as a fundamental
question posed on the research in Physics. This situation gives
relevance to the close examination of  possible mechanisms that
could be playing a role in this problem.

  Specifically, a modified perturbative QCD, altered in a way that
incorporates the presence of a condensate of zero-momenta gluons
and quarks in the initial state used for constructing the Wick
expansion, have been considered  in Refs. \onlinecite{mpla,
prd,epjc,hoyer,hoyer1,hoyer2,jhep}. An interesting aspect of these
works was the BCS-like modification of the gluon state studied in
Refs. \onlinecite{mpla, prd,epjc} which directly led to the
prediction of the constituent masses for light quarks, after
simply fixing to its currently estimated value, a relatively
unrelated quantity, the gluon condensate parameter.  Therefore, it
looked reasonable to expect that a similarly constructed state for
the quarks, after to be introduced in the treatment of massless
QCD, could have the chance of generating the Lagrangian quark
masses. The result of the preliminary consideration of this idea
in Ref. \onlinecite{jhep}, was positive. A modification of the
free-quark propagator by introducing the zero momentum terms
representing analogs of the Cooper pair condensates within this
problem was done. After that, the simplest approximation for the
Dyson equation for quarks produced a diagonal Lagrangian mass
matrix by the choice of an also diagonal structure, for the quark
condensates. Henceforth, the results of the work
\onlinecite{jhep}, pointed out that the presence of colorless
quark anti-quark condensates in the vacuum of the free theory
generating the Wick expansion, is able to produce the observed
quark mass matrix. Therefore, the question arose about the
possibility for the generation of the necessary pattern of quark
condensates by a dynamical breaking of the flavor and chiral
symmetries. A preliminary step in the study of this issue was also
done in Ref. \cite{jhep}, where the one loop contribution to the
Cornwall-Jackiw-Tomboulis (CJT) effective potential for composite
operators \cite{cjt} was evaluated as a function of the parameters
determining the quark and gluon condensates. However, it should be
underlined  that as the CJT potential is not directly giving the
ground state energy at its minimum \cite{cjt, mirans}, the
implications of the results of the evaluation for the dynamical
symmetry breaking problem under study were not clear.
 We estimate that in the case that a similar mechanism to the one
acting in standard superconductivity, could be playing a role in
the problem, the generation of large quark condensates could be
expected to be produced by strong binding color forces,  linked
with the interaction vertices. On another hand, the inability of
the standard perturbation expansion in evidencing this effect,
looks for us  to be rooted in the explicit disregarding of the
inclusion zero momentum quarks and gluons in the free vacuum
state, before the adiabatic connection of the interaction. The
incorporation of these modified Lorentz invariant vacuums in a BCS
style,  allows to produce non-trivial modifications in the Wick
expansion as it was argued in Ref. \onlinecite{prd}.

 In the former works on the theme \cite{mpla,
prd,epjc,hoyer,hoyer1,hoyer2,jhep} a troublesome aspect was
remaining about the appearance of singularities in the Feynman
diagrams due to the presence of Dirac Delta terms in the free
propagators.  This issue will be approached here in one of the
Appendices.  The basic idea of the adopted procedure is to simply
be consistent with the dimensional regularization and to extend
the appearing $\delta (0)$  like singularities for continuous $D$
dimensions in the way early introduced by Capper and Liebbrandt
\cite{capper}. This procedure lead to the outcome that these
factors simply vanish in the  $D\rightarrow 4$  limit.  An
additional recipe is also taken for the evaluation of the
$1/(p^{2}+i\epsilon )$ factors at zero momentum. These terms also
appear due to the joining of more than $n-1$ different condensate
lines in a vertex having $n$ legs.  The rule chosen in this case
will be to just regularize the scalar field propagator at the zero
value of the 4-momentum point to be equal zero.  These points are
discussed in Appendix A.  The selected prescriptions, in addition,
allow to identify the propagators evaluated in \cite{jhep}, as
modified tree propagators. They also have zero order in the
coupling $g$ series expansion after considering as the independent
parameters of the theory, the proper $g$,  in addition to the
gluon and quark condensate parameters multiplied by $g^{2}.$ This
transformation seemingly will allow to rearrange the full loop
expansion to a form in which the propagators derived in Ref.
\onlinecite{jhep} will play the role of new tree propagators.
However, this more formal aspects will be relegated for a further
study.  Here we only consider the one and two loop cases in which
it is clear that the summation over the zero order in the coupling
self-energy insertions can be done. The full demonstration of the
coincidence of the terms evaluated here with the exact loop
expansion terms of the alternative series needs only for the
checking about wether the combinatorial factors of all the lower
loop terms arising from the higher loop ones could obstacle the
proof. The verification of this issue will be considered
elsewhere.

 After giving the procedure for defining the perturbative
expansion, we continue in this work  the study of the possibility
for the dynamical generation of masses in massless QCD. Zero and
one loop vacuum contributions to the  effective potential  are
evaluated and also the full dependence on the quark condensate of
a particular two loop contribution is presented. As noticed above,
the usual definition of the effective potential was evaluated here
(\cite{mirans}) in order to avoid the non direct bounded from
below property of the directly evaluated  CJT potential
\cite{cjt}. The approximation considered consists in inserting all
the condensate dependent parts of the one loop self-energy
corrections into the free propagators, in the usual zero and one
loop correction, as follows from the discussion in Appendix A.
This procedure results in employing in the usual loop diagrams,
the propagators that produced the constituent masses for light
fermions in the work \cite{epjc},  by also considering propagators
associated to the gluon and quark condensates.  The quark
condensate dependence of only one particular two loop diagram, was
also evaluated to estimate the possible effects of the next
corrections. The evaluation of the full two loop dependence on the
condensates, in order to determine their net effect on the results
will be considered in further studies. We prefer to postpone the
calculation of these terms, waiting  for a precise definition of
the renormalization scheme in the modified expansion.

   The results obtained here for the effective potential  indicate
a dynamical generation of quark and gluon condensates.  The
dependence on the potential on the quark one,  get an unbounded
from below behavior, in the present approximation. This
instability becomes stronger by increasing the gluon condensate.
This point suggests the relevance of the presence of the gluon
condensate for the dynamical generation of the quark one.  The
unbounded from below dependence,  then indicates the need of
higher approximations for producing an eventual minimum of the
potential. The picture arising  precisely reproduce the one
expected to occur in  Ref. \cite{jhep} as a possible consequence
of the underlined analogies between the construction of the
modified wave functions for QCD and the BCS states
\cite{prd,jhep}.   The dependence of the potential on the gluon
condensate starting at zero value, is a decreasing one,  which
become steeper when the quark condensate grows.  Therefore, this
property generate the expectation about that, the yet to be
determined stabilizing contribution to the potential naturally
could show opposite behavior in the quark and gluon dependence,
able  to produce not only the stabilization, but also a value for
$M$ at the minimum laying near the known constituent mass $M=333$
$MeV$.
   At low values
of the quark condensate parameter, the gluon one develops a
minimum. However, in the present approximation, for fixed values
of the quark condensate  the value of the gluon condensate at the
minimum tend to grow, when the value of $X$ increases.  This is
not unexpected, since  we have not evaluated the full two loop
dependence on the gluon condensate, nor the possible existing
stabilizing terms. They as mentioned before, could help in
maintaining the gluon condensate at a minimum being near low
values, allowed to be fixed to the observable value. This fixation
could be done,  let say by selecting scale parameter $\mu .$  It
will be clearly surprising that the magnitude of the $X$ required
for the stabilization could  be as high as of the order of
hundreds, as it would be needed for predicting the top condensate
mass near $m_{top}=X=175$ GeV (as given by the pole of the quark
propagator for high $X$ values \cite{jhep}). However, we could not
yet disregard this possibility and the search for an  estimation
of the stabilizing terms will be undertaken.

A complementary dependence of the two loop potential as a function
of the quark condensate has been evaluated in the form of a simple
2D integral depending on this condensate and the gluon condensate
one. The outcome for this quantity turned to be finite after
including the corresponding quark condensate dependent part of the
usual quark counterterm.  This result gave us confidence about
that the  renormalization procedure can be well implemented in the
modified theory. However, a careful discussion of this question
should be considered in detail.

The introduction of the gauge parameter dependence is another
problem  which need an additional careful consideration. Partial
argues about the gauge invariance of the scheme have been done
\cite{hoyer1,hoyer2},  and also the evaluation of the one loop
gluon self-energy directly satisfied the transversality Ward
identity \cite{epjc}. The discussion in the present work and in
Ref. \cite{hoyer2} \ make also clear that the modified theory is a
multi-parameter one in which the implementation of the gauge
invariance can show subtleties needing to be carefully addressed.
However in the appendices we present some ideas about
 how consider this questions. We also expect to be able of
 presenting  more definitive conclusions in next studies.

   The work is organized as follows: in Section 2, the propagators employed
in the calculations are presented and the effective action vacuum
diagrams described. The zero and one-loop potential contributions
are discussed in Section 3.  Section 4 consider the quark
condensate dependent part of the two loop effective action
contribution being considered. The Section 5 is devoted to expose
and discuss the results of the evaluations done. Appendix A,
exposes the procedure for eliminating the singularities in the
diagram expansion through employment of dimensional regularization
and Appendix B discuss gauge invariance aspects and the
perspectives of its implementation in the proposed scheme.  In the
summary the main results of the work are shortly reviewed

\section{ Propagators and Effective Action}

In next sections the evaluation of the effective potential
including zero, one and two loop corrections will be considered.
The contributions will be calculated  by inserting the infinite
ladder of condensate dependent one-loop self-energy parts in the
original free propagators following the rules defined in Appendix
A and Ref. \cite{jhep}. These propagators for quarks and gluons,
as well as for the condensate lines (defined in Appendix A and
Ref. \cite{jhep}) are given as
\begin{widetext}
\begin{align}
G_{g\mu \nu }^{ab}(p,m)& =\frac{\delta ^{ab}}{(p^{2}-m^{2}+i\epsilon )}%
(g_{\mu \nu }-\frac{p_{\mu }\text{ }p_{\nu }}{p^{2}+i\epsilon })+\frac{\beta
\text{ }p_{\mu }\text{ \ }p_{\nu }\ }{(p^{2}+i\epsilon )^{2}}\delta ^{ab},
\label{gluon} \\
& =\frac{\delta ^{ab}}{(p^{2}-m^{2}+i\epsilon )}(g_{\mu \nu }-\frac{\beta
\text{ }p_{\mu }\text{ }p_{\nu }\ m^{2}}{(p^{2}+i\epsilon )^{2}}),\text{ \ }%
\beta =1,  \notag \\
G_{q}^{f_{1}f_{2}}(p,M,S)& =\frac{\delta ^{f_{1}f_{2}}}{\left( -p_{\mu
}\gamma ^{\mu }(1-\frac{M^{2}}{p^{2}})\ \ -\frac{S_{f_{1}}}{p^{2}}\ \right) }%
,  \label{quark} \\
\chi ^{ab}(p)& =-\frac{\delta ^{ab}}{p^{2}+i\epsilon },  \label{ghost} \\
G_{m}^{ab}& =-\frac{im^{2}}{g^{2}}\delta ^{ab}\ \delta (p),
\label{gluoncon}
\\
G_{S}& =\frac{i4\pi ^{4}S_{f}}{g^{2}C_{F}}\delta ^{ab}\delta
^{f_{1}f_{2}}\delta (p),  \label{quarkcon}
\end{align}
\end{widetext}
where(\ref{gluon}-\ref{quarkcon}) are the gluon, quark and ghost
propagators respectively and (\ref{gluoncon}), (\ref{quarkcon})
the gluon and quark condensate ones. In this work we will adopt
the general conventions for the spinor, color and Lorentz groups,
the free propagators and interactions vertices of the reference
\cite{muta}.

The parameters $m^{2},M,S_{f}$  are related with the constants $C$ and $%
C_{f}$ \ (See Ref. \cite{jhep}) characterizing the gluon and quark
condensates, as follows
\begin{align}
-m^{2}& =m_{g}^{2}=\frac{6g^{2}C}{(2\pi )^{4}},  \label{massglu} \\
S_{f}& =\frac{g^{2}C_{F}}{4\pi ^{4}}\ C_{f},  \label{S} \\
m^{2}& =f\text{ }M^{2},\text{ \ \ }f=\pm (\frac{3}{2})^{2},  \label{mM} \\
g^{2}& =g_{o}^{2}\text{ \ }\mu ^{2-\frac{D}{2}},  \label{gD} \\
D& =4-2\epsilon .  \notag
\end{align}

In these relations the parameter $f=\pm (\frac{3}{2})^{2}$ will be
considered for the two values of its sign.  However, the negative
value was the only one implied by the calculation done in
\cite{jhep}. In that work, it followed that if the parameter $C$
is chosen as positive, then the
gluon mass in the simplest approximation turns to be tachyonic $(m^{2}=-%
\frac{6g^{2}C}{(2\pi )^{4}})$ and on another hand the constituent
mass for light quarks $M$ becomes real.  The situation reverses
for negative value of the gluon parameter $C.$ However, it should
be taken into account that these relations appeared due to the
special fact that the condensate dependent part of the one gluon
self-energy had no contribution from the fermions. This situation
occurred precisely because they were assumed as massless.
Therefore, as the very same motivation of the present discussion
is related with the possibility of generating masses for the
quarks, and also paying attention to the fact that the corrected
propagators could be perhaps also constructed in a self-consistent
manner (allowing for a self-consistently generated mass), we
consider of interest to also examine the calculation for the
positive choice of  $f$ . It should be mentioned that for this
value of the parameter,  both the gluon mass and the constituent
masses are real.  Moreover, the gluon mass is near the value
estimated though other studies \cite{cornwall1}  $m=0.5$ $GeV,$
while the constituent mass gets  the reasonable value $ M=0.33$ $%
GeV$\cite{jhep}$.$  However, a study of the physical justification
of the positive choice of $f$  needs to be done.  This question
could be considered elsewhere.

It should be explicitly stated that only one flavor was assumed to
be condensed en the present discussion. This condition was elected
because at present level of approximation , the consideration of
various flavors will simply lead to an of addition identical
fermions contributions to the potential.  The question about the
possible interference of various quark condensates, since it needs
for higher order approximations for its appearance, will be
relegated to further studies.  It is clear that this is a relevant
point, because only  in the case that the presence of various
kinds of such condensates will be rejected by the system the
dynamical generation of only one (main) quark condensate will be
preferred .  This could occurs by example,  due to the presence of
terms in the effective potential growing in value when more than
one condensate are present.  This effect seems clearly possible to
occur but its considerations need for at least three loop
corrections (or their descendants according to the reasoning in
Appendix A) in which different quark loops can start to appear
\cite{jhep}.
\begin{figure}[ptb]
\begin{center}
\epsfig{figure=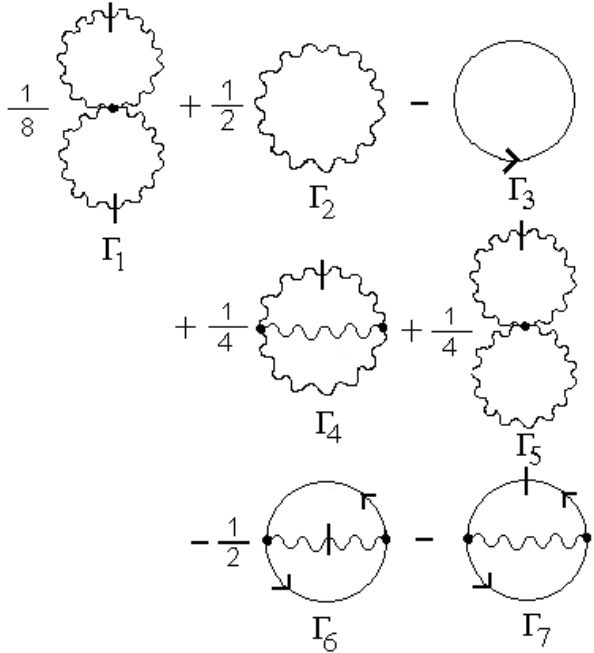,width=10cm} \vspace{-2cm}
\end{center}
\caption{ The figure shows the seven Feynman diagrams
 defining the zero and one loop contributions evaluated in the work.
 The lines having cuts
 correspond to the condensate propagators.  Therefore, although the
 associated diagrams may look as two loop ones the Delta functions
 associated to them effectively cancel one of the loop integrals. } \label{fig1}
\end{figure}

 The collection of zero,  and one loop diagrams which were evaluated \ are
illustrated in Fig.\ref{fig1}.  The diagram $\Gamma_{1}$ is the
only non having closed loops, that is  a tree correction. $ \Gamma
_{2}$ and $\Gamma _{3}$ show the usual  gluon  and quark one loop
corrections associated to the propagators  (\ref{gluon}) and
(\ref{quark}) respectively.  Further, diagrams $\Gamma_{4}$,
$\Gamma_{5},\Gamma _{6}$ and $\Gamma _{7}$ are related  with the
one loop corrections being ''descendant'' from the two loop ones
due to the cancellation  of one of the two loop integrals by a
condensate propagator,  and the insertion of all coupling $g$
independent self-energy insertions leading to the propagators
\ref{gluon}, \ref{quark} in the other two lines (See Appendix A).

\begin{figure}[ptb]
\begin{center}
\epsfig{figure=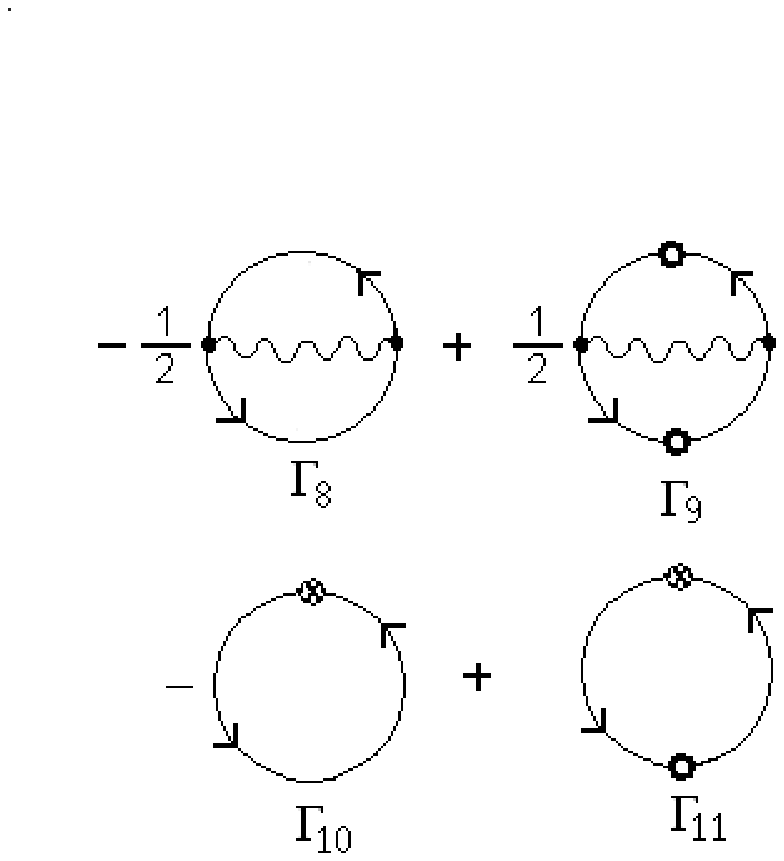,width=10cm} \vspace{-3.1cm}
\end{center}
\caption{ The figure illustrates: a) The two fermion contribution
$\Gamma_8$  considered in the work for getting a sense of the
influence of the higher loops; b) The substracted diagram
$\Gamma_9$ which is the same $\Gamma_8$ evaluated at $S=0$; c) The
diagram associated to the fermion counterterm $\Gamma_{10}$ and d)
This last diagram taken  at $S=0$ indicated by $\Gamma_{11}.$ \ }
\label{fig2}
\end{figure}
Finally in Fig. \ref{fig2} the diagrams  $\Gamma_{8}$ and $\Gamma_{9}$, $%
\Gamma_{10}$ and $\Gamma_{11}$ are defined as follows:  $ \Gamma
_{8}$ is the standard diagram for the two loop correction
including all the coupling independent self-energy insertions in
its internal lines, $\Gamma_{9}$ is the same contribution as
$\Gamma_{8}$ but taken in the limit $S\rightarrow 0$ which is
substracted in order to consider only the quark condensate
dependent part of the two loop term, the limit $S \rightarrow 0$
is indicated by the rings in the fermion lines. Finally
$\Gamma_{10}$ \ is the $g^{2} $ contribution associated to the
fermion counterterm and $\Gamma_{11}$ is the same contribution in
the limit $S\rightarrow 0$ substracted in order to again only
consider the quark condensate dependent part of the potential.
  The momentum integrations, in writing the
diagram expressions, will be taken in Minkowski space for after
perform the Wick rotation. However, it should be made precise that
in order to make the rotation without encountering poles, the sign
of $m^{2}$ should be positive ($f=(\frac{3}{2})^{2}$). However, we
will perform the Wick rotation for the two signs of $f$ without
including the terms that could be incorporated by rounding the
poles in the $p_{o}$ variables when deforming the integration
contour if $f$  takes its negative value. Therefore the results
obtained for $f$ negative, should be interpreted as the evaluation
of the effective potential  in Euclidean field theory. That is,
the evaluated quantity corresponds the thermodynamical effective
potential in the limit of zero temperature.

The employed expression for the fermion renormalization constant is given by
\cite{muta}
\begin{eqnarray*}
(Z_{2}-1) &=&-\frac{g^{2}C_{F}}{(4\pi )^{2}}\frac{\beta }{\epsilon }, \\
T_{R} &=&\frac{1}{2},\ \ C_{G}=N,\text{ \ }C_{F}=\frac{N^{2}-1}{N}.
\end{eqnarray*}

As it was remarked before, only one quark flavor will be
considered for the present qualitative discussion, since up to
this level all the quark flavors will produce a sum of
contributions all of the same functional dependence on their
respective condensates.  However, the fact that light quarks exist
furnish a guiding principle in the sense that the mass acquired by
them, if their quark condensates do not develop, should coincide
with the parameter $M$  \cite{jhep} at the minimum of the
effective potential.

The value of the renormalized gauge parameter $\beta $  is equal
to one  in accordance with the fact that the  modification of the
Feynman rules induced by the presence of the condensates was
obtained in \cite{prd} within the Feynman  gauge $\beta =1.$ As it
was mentioned in the Introduction, the question of the gauge
parameter  invariant formulation of the modified perturbation
theory under study, which clearly represent a required step in the
formal completion of the approach should be further considered.
However, some general remarks on this question are given in
Appendix B and a more concrete study is expected to be considered
elsewhere.
\section{zero and one-loop terms}

In this section the results for the evaluation of contribution to
the one loop effective potential $\Gamma_{1}$ to $\Gamma_{7}$ will
exposed below in consecutive order.

\subsection{Zero loop term}

The direct substitution \ of the gluon condensate propagator
(\ref{gluoncon}) in the analytic expression associated to
$\Gamma_{1}$, after evaluating all the Lorentz, spinor and color
traces leads to
\begin{equation*}
\Gamma ^{(0)}=-\frac{2m^{4}}{g^{2}}=-V^{(0)}.
\end{equation*}
That is, a positive potential proportional to $m^{4}$.  As the one
loop terms have zero order in the coupling $g,$ in the expansion
in powers of the parameters defined in Appendix A, this term shows
a  power -2 of $g$,  since the original diagram was of order two
and there is two condensate lines in the diagram which reduce the
power in four according to (\ref{power}). \medskip
\subsection{Standard one loop terms }

The sum of the one loop terms corresponding to $\Gamma _{2}$ and $
\Gamma _{3}$ in Fig.\ref{fig1} have the form
\begin{widetext}
\begin{eqnarray*}
\Gamma _{gf}^{(1)} &=&\Gamma _{g}^{(1)}+\Gamma _{f}^{(1)}+\Gamma
_{S}^{(1)}=-V_{g}^{(1)}-V_{f}^{(1)}-V_{S}^{(1)} \\
&=&-\frac{i}{2}\text{ }Tr\text{ }\left[ \log \left[ G_{g}^{-1}(0)\text{ }%
G_{g}^{-1}(m)\right] \right] +i\text{ }Tr\text{ }\left[ \log \left[
G_{q}^{-1}(0,0)\text{ }G_{q}^{-1}(M,0)\right] \right] + \\
&&+i\text{ }Tr\text{ }\left[ \log \left[ G_{q}^{-1}(0,0)\text{ }%
G_{q}^{-1}(M,S)\right] \right] -i\text{ }Tr\text{ }\left[ \log \left[
G_{q}^{-1}(0,0)\text{ }G_{q}^{-1}(M,0)\right] \right] ,
\end{eqnarray*}
\end{widetext}
which have been expressed as the sum of a $S$ independent term
corresponding to the same diagrams evaluated at $S=0$  plus a $S$
dependent contribution vanishing  in the limit $S\rightarrow 0.$
 After calculating the Lorentz, spinor and color traces for the
$S=0$ gluon and quark  loops $\Gamma _{g}^{(1)}$ and  $\Gamma
_{f}^{(1)},$ and dimensionally regularizing the integral,  it
follows
\begin{eqnarray*}
\Gamma _{g}^{(1)} &=&-\frac{(N^{2}-1)(D-1)}{2}\int \frac{dp^{D}}{(2\pi )^{D}i%
}\log \left[ \frac{p^{2}}{p^{2}-m^{2}}\right] , \\
\Gamma _{f}^{(1)} &=&4N\int \frac{dp^{D}}{(2\pi )^{D}i}\log \left[ \frac{%
p^{2}}{p^{2}-M^{2}}\right] .
\end{eqnarray*}

\ But, in both cases the recourse of taking the derivative of  the
expressions over the parameters  in the gluon and quark cases
leads to simpler expressions. Then, after also performing the Wick
rotation in the temporal momentum component according to
\begin{equation*}
p_{0}\rightarrow i\text{ }p_{4},
\end{equation*}
the derivative  over the parameters expressions can be integrated
in momentum space by employing the formula \cite{muta}
\begin{equation*}
\int_{E}\frac{dp^{D}}{(2\pi )^{D}}\left[ \frac{1}{p^{2}+L^{2}}\right] =\frac{%
B(\frac{D}{2},1-\frac{D}{2})}{(4\pi )^{\frac{D}{2}-2}\Gamma (\frac{D}{2})}%
(L^{2})^{D-2},
\end{equation*}
in which $L$ can be selected as  $m$ or $M$ as appropriate for the
gluon or quark terms respectively.  The results of the integrals
after to be integrated over the parameter again from their zero
values to the original ones, by also considering that the extended
dimension $D$ is such the real part of $D-2$  is positive, allows
to obtain

\begin{eqnarray}
\Gamma _{g}^{(1)}(m) &=&\frac{(D-1)(N^{2}-1)B(\frac{D}{2},1-\frac{D}{2})
 (m^{2})^{\frac{D}{2%
}} }{D%
\text{ }(4\pi )^{\frac{D}{2}}\Gamma (\frac{D}{2})},\text{ }  \label{1loopglue} \\
\Gamma _{q}^{(1)}(M) &=&-\text{ }\frac{8\text{ }N\text{ }B(\frac{D}{2},1-%
\frac{D}{2})(M^{2})^{\frac{D}{2}}}{D\text{ }(4\pi )^{\frac{D}{2}}
\Gamma (\frac{D}{2})}.\text{ }%
,  \label{1loopquark}
\end{eqnarray}

\ At this point after considering relations (\ref{massglu}),
(\ref{mM}) and (\ref{gD}) defining $m$ and $\ M$ as functions of
the dimension $D$ , and substracting the pole part in $\epsilon $
of (\ref{1loopglue}) and (\ref {1loopquark}), the Minimal
Substraction result for the one loop effective action  is
\begin{widetext}
\begin{eqnarray}
V_{g}^{(1)}(m) &=&-\frac{\text{ }(N^{2}-1)}{128\text{ }\pi ^{2}}\text{ }%
m^{4}\left( -6\log \left( \frac{m^{2}}{4\pi \mu ^{2}}\right) -6\gamma
+5)\right) , \\
V_{q}^{(1)}(M) &=&\frac{3\text{ }(N^{2}-1)}{128\text{ }\pi ^{2}}\text{ \ }%
M^{4}\left( -2\log \left( \frac{M^{2}}{4\pi \mu ^{2}}\right)
-2\gamma +3)\right) .
\end{eqnarray}

The $S$ dependent correction $\Gamma _{S}^{(1)}$ after all the trace
evaluations can be written as
\begin{eqnarray*}
\Gamma _{q,S}^{(1)} &=&+i\text{ }Tr\text{ }\left[ \log \left[ G_{q}^{-1}(0,0)%
\text{ }G_{q}^{-1}(M,S)\right] \right] - \\
& &-i\text{ }Tr\text{ }\left[ \log \left[
G_{q}^{-1}(0,0)\text{ }G_{q}^{-1}(M,0)\right], \right]  \\
&=&2N\int \frac{dp^{D}}{(2\pi )^{D}i}\log \left[ \frac{p^{2}(p^{2}-M^{2})^{2}%
}{p^{2}(p^{2}-M^{2})^{2}-S^{2}}\right] .
\end{eqnarray*}

This integral after the Wick rotation is convergent in the limit $%
D\rightarrow 4$ and takes the form
\begin{eqnarray*}
\Gamma _{q,S}^{(1)} &=&-V_{q,S}^{(1)}=-2\text{ }N\int_{E}\frac{dp^{D}}{(2\pi
)^{D}}\log \left[ \frac{p^{2}(p^{2}+M^{2})^{2}}{p^{2}(p^{2}+M^{2})^{2}+S^{2}}%
\right],  \\
&=&-4\text{ }N\text{ }\int_{0}^{\infty }\frac{dp\text{ }p^{3}}{(4\pi )^{2}}%
\text{ }\log \left[ \frac{p^{2}(p^{2}+M^{2})^{2}}{%
p^{2}(p^{2}+M^{2})^{2}+S^{2}}\right] .
\end{eqnarray*}

\subsection{One loop terms descending from the two loop gluon diagrams}
  After writing  the analytical expressions for the diagram $\Gamma
_{4}$  and evaluating the Lorentz, spinor and color traces, the
expression can be rewrote in the form

\begin{eqnarray*}
\Gamma _{2g}^{(1,1)} &=&-V_{2g}^{(1,1)}=\frac{(N^{2}-1)N\ m^{2}}{4(2\pi )^{D}%
}\int \frac{dp^{D}}{(2\pi )^{D}i}\frac{(-6\ p^{2}D+2(D+11)\beta m^{2}-8\
m^{4}/p^{2})}{(p^{2}-m^{2})^{2}}, \\
&=&\frac{(N^{2}-1)N\ m^{2}}{4(2\pi )^{D}}\int_{E}\frac{dp^{D}}{(2\pi )^{D}}%
\frac{(6\ p^{2}D+2(D+11)\beta m^{2}+8\ m^{4}/p^{2})}{(p^{2}+m^{2})^{2}}.
\end{eqnarray*}

In the second line of this equation the Wick rotation has been
done. The integrals can be explicitly performed to give

\begin{eqnarray*}
\Gamma _{2g}^{(1,1)} &=&G1=\frac{(N-1)N\ \Gamma (1-\frac{D}{2})\left[
3D^{2}+2\beta (D+11)(1-\frac{D}{2})\right] }{4(2\pi )^{D}(4\pi )^{\frac{D}{2}%
}}(m^{2})^{\frac{D}{2}}+ \\
&&+\frac{2(N-1)N\ \Gamma (\frac{D}{2}-1)\Gamma (3-\frac{D}{2})}{(2\pi
)^{D}(4\pi )^{\frac{D}{2}}\Gamma (\frac{D}{2})}(m^{2})^{\frac{D}{2}}.
\end{eqnarray*}

After applying the same procedure for the analytical expressions
associated to the diagram  $\Gamma _{5}$\ the result  is

\begin{eqnarray*}
\Gamma _{2g}^{(1,2)} &=&-V_{2g}^{(1,2)}=\frac{(N^{2}-1)N(D-1)\ m^{2}}{2(2\pi
)^{D}}\int \frac{dp^{D}}{(2\pi )^{D}i}\frac{1}{(p^{2}-m^{2})}\left[ 1-\frac{%
\beta m^{2}}{p^{2}}\right],  \\
&=&-\frac{(N^{2}-1)N(D-1)\ m^{2}}{4(2\pi )^{D}}\int_{E}\frac{dp^{D}}{(2\pi
)^{D}}\frac{1}{(p^{2}+m^{2})}\left[ 1+\frac{\beta m^{2}}{p^{2}}\right] , \\
&=&-\frac{(N-1)N\ (D-1)(D-\beta )\Gamma (1-\frac{D}{2})}{2(2\pi )^{D}(4\pi
)^{\frac{D}{2}}}(m^{2})^{\frac{D}{2}}.
\end{eqnarray*}
\end{widetext}
It is an interesting outcome that after adding these two
contributions and removing the dimensional regularization limit,
the result remains finite,  a fact that also delete the
logarithmic terms in the outcome. The total contribution of these
terms for the potential at the end takes the form
\begin{equation*}
\lim_{D\rightarrow 4}(V_{2g}^{(1,1)}+V_{2g}^{(1,2)})=\frac{3f^{2}M^{4}}{8\pi
^{2}}.
\end{equation*}
\begin{widetext}
\subsection{One loop terms descending  from the two-loop quark diagram}
The last one loop diagrams  $\Gamma_6 $  and $\Gamma_{7} $
correspond to the descendants of the two loop terms having a
closed fermion line.  The integral expression obtained for them
after performing the Lorentz, spinor and color traces are not so
simple and we just numerically  evaluate them in this work. The
resulting integral expressions are
\begin{eqnarray}
\Gamma _{2q}^{(1,1)}&=&-V_{2q}^{(1,1)}=\frac{(N^{2}-1)\ m^{2}}{3(4\pi )^{2}}%
\int_{0}^{\infty }dp\ p^{3}\frac{(2\ p^{2}(\ p^{2}+M^{2})^{2}+D\ S^{2})}{%
(p^{2}(\ p^{2}+M^{2})^{2}+S^{2})^{2}}, \nonumber \\
\Gamma _{2q}^{(1,2)}&=&-V_{2q}^{(1,2)}=-N\ S^{2}\int_{E}\frac{dp}{(2\pi )^{4}}%
\ \frac{(D\ p^{2}+m^{2}-i\epsilon )}{(p^{2}(\ p^{2}+M^{2})^{2}+S^{2})(\
p^{2}+m^{2}-i\epsilon )}.
\end{eqnarray}
In ending this section it can be remarked the interesting outcome
that all the ''descendant'' diagrams  became finite ones.
\end{widetext}
\section{two-loop quark term}

As we are interested in the studying the generation of the quark
condensates we will consider here only the full dependence on this
quantity  associated to the diagram  $\Gamma _{8}$ in Fig.
\ref{fig2} defined by a fermion loop formed with propagators
(\ref{quark}), showing  two quark-gluon interaction vertices.
Therefore,the same diagram expression but evaluated at $S=0$ will
be substracted from this contribution. This terms is associated
with $\Gamma _{9}$ \ in Fig. \ref{fig2}. This substraction simply
corresponds to the same analytic expression of the diagram but
taken for $S=0$ and is represented by the same figure but having
small rings in the quark propagators.  As the diagram associated
to the fermion counterterm of the standard massless QCD $\Gamma
_{10}($of order $g^{2}$ and therefore needed for renormalization
at one loop level) is also depending on the condensate parameter
$S$, the same kind of substraction is done of  the $S=0$
counterterm term associated to $\Gamma _{11},$ in which again the
ring in the quark line means the evaluation in $S=0.$

The substracted terms, exactly give the full  two loop term formed
by two quark propagators and one gluon line of the theory in the
absence of  the fermion condensate.  As mentioned before we will
relegate the evaluation of the full two loop gluon parameter
dependence to further studies. The main reason for doing so is
that for these terms, it is more relevant to precisely define the
way in which the renormalization should be done within the
considered scheme. As it is also following in Appendix A, at the
two loop level there will appear additional quark condensate
dependence coming from two loop diagrams being descendant from
higher loop terms of the original expansion as discussed in
Appendix A.  From exploring evaluations we know however, that it
seems possible to cancel the two loop infinities by renormalizing
the condensate parameters. However, a clearer understanding on the
structure of the allowed counterterms in the modified expansion is
yet desirable before evaluating the full  two loop terms.

  After evaluating the spinor and color traces in the analytic
expressions corresponding to the Feynman graphs appearing in
Fig.\ref {fig1}, the considered contributions to the effective
potential can be written in the form
\begin{widetext}
\begin{eqnarray}
\Gamma _{fg}^{(2)}(M,S) &=&-\frac{(N^{2}-1)g^{2}}{4}\int \frac{d^{D}q}{(2\pi
)^{D}i}\frac{d^{D}q^{\prime }}{(2\pi )^{D}i}\frac{1}{((q-q^{\prime
})^{2}-m^{2})}\times   \label{gf} \\
&&\frac{1}{(q^{2}(q^{2}-M^{2})^{2}-S^{2})(q^{\prime 2}(q^{\prime
2}-M^{2})^{2}-S^{2})}\times   \notag \\
&&{\LARGE \{}-4q^{2}q^{\prime 2}(q^{2}-M^{2})(q^{\prime 2}-M^{2}){\Large [}%
(D-2)q.q^{\prime }-\beta \frac{m^{2}}{(q^{\prime }-q)^{2}}\times   \notag \\
&&(q.q^{\prime }-\frac{2q.(q^{\prime }-q)q^{\prime }.(q^{\prime }-q)}{%
(q^{\prime }-q)^{2}}){\Large ]}+4(D-\frac{\beta \text{ }m^{2}S^{2}}{%
(q^{\prime }-q)^{2}})\text{ }q^{2}q^{\prime 2}{\LARGE \},}  \notag \\
\Gamma _{fg}^{(2)}(M,0) &=&-\frac{(N^{2}-1)g^{2}}{4}\int
\frac{d^{D}q}{(2\pi )^{D}i}\frac{d^{D}q^{\prime }}{(2\pi
)^{D}i}\frac{(-4)}{((q-q^{\prime
})^{2}-m^{2})(q^{2}-M^{2})(q^{\prime 2}-M^{2})}\times   \label{gfsubs} \\
&&{\Large [}(D-2)q.q^{\prime }-\beta \frac{m^{2}}{(q^{\prime }-q)^{2}}%
(q.q^{\prime }-\frac{2q.(q^{\prime }-q)q^{\prime }.(q^{\prime }-q)}{%
(q^{\prime }-q)^{2}}){\Large ]}{\LARGE ,}  \notag \\
\Gamma _{fC}^{(2)}(M,S) &=&4N(Z_{2}-1)\int \frac{d^{D}q}{(2\pi )^{D}i}\frac{%
(q^{2})^{2}(q^{2}-M^{2})}{q^{2}(q^{2}-M^{2})^{2}-S^{2}},  \label{count} \\
\Gamma _{fC}^{(2)}(M,0) &=&4N(Z_{2}-1)\int \frac{d^{D}q}{(2\pi )^{D}i}\frac{%
q^{2}}{(q^{2}-M^{2})}.  \label{countsubs}
\end{eqnarray}
\end{widetext}
It can be noticed  that mass dimension of the parameter $S$ is
equal to 3, that is a relatively high value. Therefore, the terms
of the expansion in powers of $S$  for the denominator of the
integrand associated to $\Gamma _{gf}^{(2)}$ will have three
powers of the momentum convergence factors for each power of $S$
appearing in the expansion.  The same effect is occurring in the
fermion counterterm $\Gamma _{fC}^{(2)}.$

Then, it follows that the quantity
\begin{widetext}
\begin{equation*}
\Gamma _{fg}(m,M,S,\epsilon )=\Gamma _{gf}^{(2)}(M,S,\epsilon )-\Gamma
_{gf}^{(2)}(M,0,\epsilon )+\Gamma _{fC}^{(2)}(M,S,\epsilon )-\Gamma
_{fC}^{(2)}(M,0,\epsilon ),
\end{equation*}
\end{widetext}
which contains, by construction, the whole dependence of the
effective action on the fermion condensate parameter $S, $ turns
to be finite in the limit $D \rightarrow 4$   $(\epsilon
\rightarrow 0).$ This result is simply expressing the fact that
the renormalization constant $Z_{2}$ of the massless QCD
(determined in the absence of any condensate) is also able to
extract the infinities from the single fermion condensate
dependent contribution under study.  As noticed before, according
to the above described substraction procedure, the substracted
terms in addition with the non considered two loop ones, exactly
correspond to the two loop plus counterterm contributions in the
absence of the fermion condensate. This terms, including the ones
descending form the higher loops (according to the reasons given
in Appendix A) will not be considered here.

  The finite contribution $\Gamma _{fg}$ before passing to
Euclidean variables can be written as the sum  of the following
three terms
\begin{widetext}
\begin{eqnarray}
\Gamma _{fg}^{(2)} &=&\Gamma _{fg}^{(2,1)}+\Gamma _{fg}^{(2,2)}+\Gamma
_{fg}^{(2,3)} \\
\Gamma _{fg}^{(2,1)} &=&\frac{(N^{2}-1)g^{2}}{2}\int \frac{d^{D}q}{(2\pi
)^{D}i}\int {\LARGE (}\frac{d^{D}q^{\prime }}{(2\pi )^{D}i}\frac{4\text{ }%
(D-2)\text{ }q.q^{\prime }}{((q-q^{\prime })^{2}-m^{2})(q^{\prime 2}-M^{2})}-
\\
&&-\frac{4}{(4\pi )^{2}}\frac{q^{2}}{{\Large \epsilon }}{\LARGE )}\times
\frac{S^{2}}{(q^{2}(q^{2}-M^{2})^{2}-S^{2})(q^{2}-M^{2})}+  \notag \\
&&(N^{2}-1)g^{2}\int \frac{d^{D}q}{(2\pi )^{D}i}\frac{d^{D}q^{\prime }}{%
(2\pi )^{D}i}\times \frac{2\text{ }(D-2)q.q^{\prime }S^{4}}{%
(q^{2}(q^{2}-M^{2})^{2}-S^{2})}\times   \notag \\
&&\frac{1}{(q^{\prime 2}(q^{\prime
2}-M^{2})^{2}-S^{2})(q^{2}-M^{2})(q^{\prime 2}-M^{2})((q-q^{\prime
})^{2}-m^{2})},  \notag \\
\Gamma _{fg}^{(2,2)} &=&-\frac{\beta \text{ }m^{2}(N^{2}-1)g^{2}}{4}\int
\frac{d^{D}q}{(2\pi )^{D}i}\frac{d^{D}q^{\prime }}{(2\pi )^{D}i}\times \frac{%
4\text{ }q^{2}q^{\prime 2}(q^{2}-M^{2})}{(q^{2}(q^{2}-M^{2})^{2}-S^{2})}%
\times  \\
&&\frac{(q^{\prime 2}-M^{2})(2\text{ }q^{2}q^{\prime 2}-q\mathbf{.}q^{\prime
}(q^{2}+q^{\prime 2}))}{(q^{\prime 2}(q^{\prime
2}-M^{2})^{2}-S^{2})((q-q^{\prime })^{2}-m^{2})((q^{\prime }-q)^{2})^{2}},
\notag \\
\Gamma _{fg}^{(2,3)} &=&\frac{\beta \text{ }m^{2}(N^{2}-1)g^{2}}{4}\int
\frac{d^{D}q}{(2\pi )^{D}i}\frac{d^{D}q^{\prime }}{(2\pi )^{D}i}\times \frac{%
4\text{ }q^{2}q^{\prime 2}S^{2}}{(q^{2}(q^{2}-M^{2})^{2}-S^{2})}\times  \\
&&\frac{S^{2}}{(q^{\prime 2}(q^{\prime 2}-M^{2})^{2}-S^{2})((q-q^{\prime
})^{2}-m^{2})(q^{\prime }-q)^{2}},  \notag
\end{eqnarray}
\end{widetext}
where the term showing the $\frac{1}{\epsilon }$ factor is
associated to the fermion counterterm. It is responsible for the
substraction of the divergent part of the remaining expressions.

After performing the Wick rotation,  it is possible  to eliminate
the pole term in $\epsilon$ by using the identity
\begin{equation*}
\frac{1}{\epsilon }=\frac{(4\pi )^{\frac{D}{2}}\Gamma (\frac{D}{2})\left[
-q^{2}\right] ^{2-\frac{D}{2}}}{\epsilon B(\frac{D}{2},2-\frac{D}{2})B(\frac{%
D}{2}-1,\frac{D}{2}-1)}\int \frac{d^{D}q^{\prime }}{(2\pi )^{D}i}\frac{1}{%
(q-q^{\prime })^{2}q^{\prime 2}}.
\end{equation*}
  Then,  the finite fermion condensate dependent contribution to the
particular two loop term evaluated here,  in the limit $\epsilon
\rightarrow 0,$ can be expressed as follows
\begin{eqnarray*}
V_{fg} &=&-\Gamma _{ffg}=-v_{0}\left[
v_{f}^{(1)}+v_{f}^{(2)}+v_{f}^{(3)}+v_{f}^{(4)}+v_{f}^{(5)}\right] , \\
v_{0} &=&\frac{4}{(4\pi )^{4}}(N^{2}-1)g^{2}M^{4},
\end{eqnarray*}

The quantities $v_{f}^{(i)},i=1,2,3$,4,5 appearing above were
reduced to simple 2D integrals after performing the angular
integrations in the 4-dimensional Euclidean space.  They take the
explicit forms
\begin{widetext}
\begin{eqnarray*}
v_{f}^{(1)}=-2\ X^{6}\int_{0}^{\infty }dq\int_{0}^{\infty }dq^{\prime } &&%
\frac{q^{4}q^{\prime 4}}{(q^{2}(q^{2}+1)^{2}+X^{6})(q^{\prime 2}(q^{\prime
2}+1)^{2}+X^{6})}\times  \\
&&\ln \left( \frac{q^{2}+q^{\prime 2}+2qq^{\prime }+f-i\epsilon }{%
q^{2}+q^{\prime 2}-2qq^{\prime }+f-i\epsilon }\right),
\end{eqnarray*}
\ \ \ \
\begin{eqnarray*}
v_{f}^{(2)}=+\ X^{6}\int_{0}^{\infty }dq\int_{0}^{\infty }dq^{\prime } &&%
\frac{q^{3}q^{\prime 3}}{(q^{2}(q^{2}+1)^{2}+X^{6})(q^{2}+1)}\times {\LARGE %
\{}\frac{1}{q^{\prime 2}(q^{\prime 2}+1)}+ \\
&&+\frac{q^{2}+q^{\prime 2}+f-i\epsilon }{4\text{ \ }qq^{\prime }(q^{\prime
2}+1)}\times \ln \left( \frac{q^{2}+q^{\prime 2}+2qq^{\prime }+f-i\epsilon }{%
q^{2}+q^{\prime 2}-2qq^{\prime }+f-i\epsilon }\right)  \\
&&-\frac{q^{2}+q^{\prime 2}-i\epsilon }{4\text{ }qq^{\prime 3}}\ln \left(
\frac{q^{2}+q^{\prime 2}+2qq^{\prime }-i\epsilon }{q^{2}+q^{\prime
2}-2qq^{\prime }-i\epsilon }\right) {\LARGE \}},
\end{eqnarray*}
\begin{eqnarray*}
\text{ \ \ }v_{f}^{(3)} &=&-X^{12}\int_{0}^{\infty }dq\int_{0}^{\infty
}dq^{\prime }\frac{q^{3}q^{\prime 3}}{(q^{2}(q^{2}+1)^{2}+X^{6})(q^{\prime
2}(q^{\prime 2}+1)^{2}+X^{6})(q^{2}+1)(q^{\prime 2}+1)}\times  \\
\text{ \ \ \ \ \ } &&{\LARGE \{}-1{\LARGE +}\frac{q^{2}+q^{\prime
2}-i\epsilon }{4\text{ \ }q\text{ }q^{\prime }}\ln \left( \frac{%
q^{2}+q^{\prime 2}+2qq^{\prime }-f^{2}-i\epsilon }{q^{2}+q^{\prime
2}-2qq^{\prime }-f^{2}-i\epsilon }\right) {\LARGE \}},
\end{eqnarray*}
\begin{eqnarray*}
\text{\ }v_{f}^{(4)}=-\beta \int_{0}^{\infty }dq\int_{0}^{\infty }dq^{\prime
} &&q^{3}q^{\prime 3}(q^{2}+1)(q^{\prime 2}+1)\times \text{ \ \ \ \ \ \ \ \
\ \ \ \ \ \ \ } \\
&&{\LARGE \{}\frac{q^{2}q^{\prime 2}}{(q^{2}(q^{2}+1)^{2}+X^{6})(q^{\prime
2}(q^{\prime 2}+1)^{2}+X^{6})}-\frac{1}{(q^{2}+1)^{2}(q^{\prime 2}+1)^{2}}%
{\LARGE \}}\times  \\
&&{\LARGE \{}1+\frac{(q^{2}+q^{\prime 2}-\frac{(q^{2}-q^{\prime 2})^{2}}{f}}{%
4\text{ \ }q\text{ }q^{\prime }}\ln \left( \frac{(q-q^{\prime
})^{2}+f-i\epsilon }{(q+q^{\prime })^{2}+f-i\epsilon }\frac{(q+q^{\prime
})^{2}-i\epsilon }{(q-q^{\prime })^{2}-i\epsilon }\right) {\LARGE \}},
\end{eqnarray*}
\begin{eqnarray*}
\text{ \ \ }v_{f}^{(5)}=-\frac{\beta X^{6}}{2}\int_{0}^{\infty
}dq\int_{0}^{\infty }dq^{\prime } &&\frac{q^{4}q^{\prime 4}}{%
(q^{2}(q^{2}+1)^{2}+X^{6})(q^{\prime 2}(q^{\prime 2}+1)^{2}+X^{6})}\times  \\
&&\ln \left( \frac{(q-q^{\prime })^{2}+f-i\epsilon }{(q+q^{\prime
})^{2}+f-i\epsilon }\frac{(q+q^{\prime })^{2}-i\epsilon }{(q-q^{\prime
})^{2}-i\epsilon }\right) ,
\end{eqnarray*}
\end{widetext}
in which as before the dimensionless quantities $X$  is given as
follows
\begin{equation}
S=M^{3}X^{3}.  \notag
\end{equation}
The $\epsilon $ parameter is retained here since  it helps to
regularize the integrals even in the Euclidean case when $f$ is
negative.

\section{Discussion}

\ In this section we will present the results for the evaluation
of the effective potential as a function of the condensate
parameters  $M,$ $S$ and the couplings constant $g.$  The
calculations were done for the two signs of the parameter $f$
that defines the relation between the constituent quark and gluon
mass parameters $m$ and $M$ through
\begin{equation*}
m^{2}=f\ \ M^{2}.
\end{equation*}
 As it was remarked before, only the negative sign was arising in
the work  \cite{epjc}  because the constituent  mass value
evaluated in that work was satisfying the above relation with the
negative sign. This fact was a direct consequence of the free
standard quark propagator being massless. However, as it was
noticed before here,   we suspect that a sort of self-consistent \
treatment could lead to an unrestricted sign of $f$ . Thus the
evaluation for positive $ f $ values was also considered. An
interesting point in this sense, is that for positive $f$  the
results for the potential are completely real, a fact that is not
occurring for the  more relevant case under study, that is
$f=-(\frac{3}{2})^{2}$.  Moreover, in this situation, it turned
out that the value of $m$ following,  once the quark condensate
$<g^{2}G^{2}>$ is fixed, is $m=0.5$ $GeV$ , which coincides with a
result obtained in Ref. \cite{cornwall1}. Unfortunately, the sign
of the light quark masses which follows from the Dyson equation
\cite{epjc} is opposite to the sign of the gluon condensate
parameter, that is basically the sign of $m^{2}.$ Therefore, in
case that we select the negative sign of $\ f$ , then the absolute
value of the constituent mass for light quarks will be also $333$
$MeV$ but the mass will be tachyonic.  This perhaps is another
possibility which could be needed to be also examined. In any
case, neither gluons or quarks appear in Nature and  perhaps both
will be absent as real excitations in both descriptions in which
none of them will be asymptotic states after including more
corrections \cite{jhep}.

 Let us define for the graphical illustrations  the quantities $%
V(X,M,g,\mu )$ \ and its imaginary part $\ V_{im}(X,M,g,\mu )$, as
the sum of all the contributions to the effective potential (the
negatives of the effective action terms) evaluated in previous
sections  divided by the constant factor  $1/(8\pi ^{4})$.  In the
various figures below the dependence of the quantity $ V$ \ or its
imaginary part $ V_{im}$ are plotted as functions of two of their
arguments selecting the others as given by  characteristic values
of interest in the present state of the discussion. The plots are
associated to the relevant case $f=-(3/2)^{2}$ and sometimes
comments about the effect of the graphs of changing the sign of
$f$  will be done.
\begin{figure}
\epsfig{figure=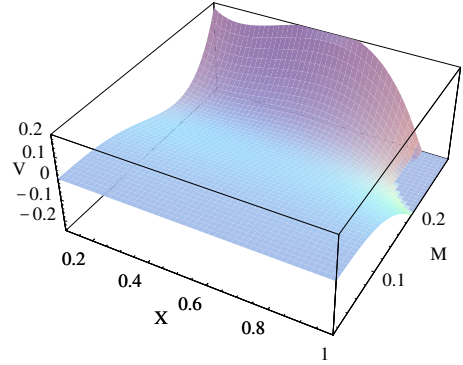,width=6cm} \vspace{-0.5cm}
\caption{ The landscape picturing the dependence of the function V
(the effective potential divided by $1/(8\pi^{4})$) on the quark
condensate (measured by $X$) and the gluon one (measured by the
constituent mass $M$). The values of the other parameters are
$g=2.74$ and $\mu=6.8 \ GeV $. At $X=0$ the potential develops a
minimum at certain value of $M$ which can be varied by changing
the scale parameter $\mu$. The dependence on $X$ indicates an
instability upon the generation of  values of $X$
 which is not controlled at large $X$ values. } \label{fig3}
\end{figure}
The  Fig. \ref{fig3} illustrates the behavior of the effective
potential as a function $X$ and the  constituent mass $M.$ both
quantities are defined in the text in terms of the quark
condensate and the gluon one through

\begin{eqnarray*}
X &=&\frac{S^{\frac{1}{3}}}{M}=\frac{1}{M}(\frac{gC_{F}}{4\pi ^{4}}C_{f})^{%
\frac{1}{3}},\\
M &=&\frac{m^{2}}{f},\text{ \ \ \ \ \ \ for }M\text{ real.}
\end{eqnarray*}
\ The value of the coupling $g$ selected for the plot was $
g=2.74$  which corresponds to a strong coupling value $\alpha
=\frac{g^{2}}{4\pi }$  being near $ 0.6.$  In addition the mass
scale parameter value $\mu =6.8\ GeV$ was fixed. Note that the
minimum at zero quark condensate $X=0$ is laying  near $200$
$MeV,$ which is lower but near the constituent mass value $M=333$
$ MeV.$ It is interesting that to fix minimum of the potential for
$X=0,$ at this value of $M$ requires a relatively large value of
$\mu.$
\begin{figure}
\begin{center}
\epsfig{figure=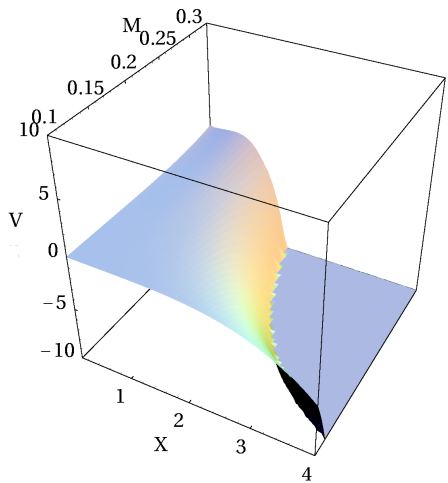,width=6cm}
\end{center}
\caption{ The dependence of the function V on $X$ and $M$ for a
wider range of values of $X$ and the same fixed parameters
$g=2.74$ and $\mu=6.8$ $GeV $. It shows more explicitly that the
system in the taken approximation does not show a minimum of the
potential when $X$ grows.  It can be also seen that even at zero
value of the gluon condensate ($M=0$) the potential remains
unbounded from below. That is, the sole presence of the quark
condensate also makes the system unstable under the generation of
$X$ from the state at $X=0$. This property suggests that the
instability effect could not be destroyed by the deconfinement
transition at high temperaures. } \label{fig4}
\end{figure}

As it can be observed, the landscape of the potential makes clear
that the system at $ X=0$ dynamically develops a gluon condensate
parameter with a potential similar in form to the Savvidi one in
the early Chromomagnetic field models. (\cite{savvidi,shabad}). It
also can be seen that the system at zero values of both parameters
shows an instability upon the generation of both gluon and quark
condensate.  The instability is stronger for the dynamical
generation of the quark condensate.  It can be also observed that
the increasing of the gluon condensate parameter makes stronger
the instability to the generation of the parameter $X.$ This
property  is supporting the expectation expressed in
\cite{epjc,jhep} about that the  color coupling could produce a
sort superconductivity effect being able to generate intensive
quark condensate values alike to the Ginzburg-Landau fields.  If
such effect is really occurring in Nature, the  Top Condensate
model could emerge as a possible effective field theory determined
by the strong forces and upon this the Higgs fields could be no
other thing as the Top condensate value. \cite{bardeen}.  This
occurrence could also explain the similarities between the
properties of the quark mass spectrum and the spectrum of
superconductivity systems, underlined in the ''Democratic Symmetry
Breaking'' analysis \cite{fritzsch}.

 In order to evidence the behavior of the potential at larger
values of $X$, in Fig \ref{fig4}, almost the same plot but for an
increased range of the variable $X$ is depicted.  The picture
clearly show that,  in the framework of the present approximation,
and for reasonable values of the coupling (\ $\alpha =0.6)$, there
are no terms that control the instability for the generation of
the quark condensate parameter, which under the shown potential
will tend to grow without limit. Therefore, it becomes clear that
the stabilization of the minimum of the system should come from
terms higher than the ones considered here. Precisely in
\cite{jhep},  this behavior was guessed to occur thanks to  the
color interaction between quarks.  Therefore, under the assumption
that the technique being used in this exploration is well
describing the massless QCD, it seems that this theory  could
dynamically develop heavy quark masses.
\begin{figure}[ptb]
\begin{center}
\epsfig{figure=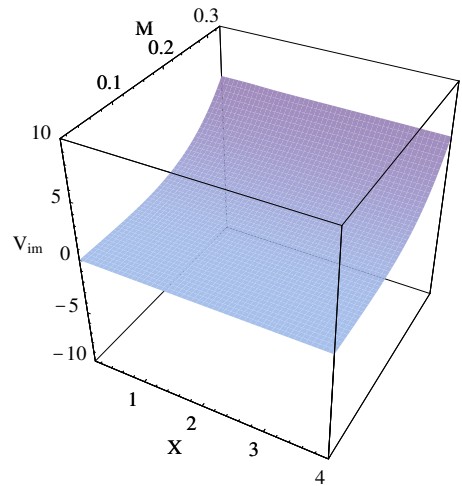,width=6cm}
\end{center}
\caption{  The plot of the function $V_{im}$ for the same range of
$X$ and $M$ used in Fig. \ref{fig4} and the same fixed parameters
$g=2.74$ and $\mu=6.8$ $GeV $ and $f=-(3/2)^2$. It can be seen
that the ratio between the imaginary and the real part of the
potential decreases for the higher values of $X$. For $f=(3/2)^2$
the potential is real.} \label{fig5}
\end{figure}
This outcome could be another  realization of the dimensional
transmutation effect \cite{coleman}. A requirement for the next
corrections to produce helpful results for modelling,  is that the
stabilizing potential at large $X$  values behaves in such a way,
that its dependence on $M$ assures that the extreme point occurs
at low values of $M$. Then, it could be expected that it can be
fixed to the observable value near $333$ $MeV$  by selecting
appropriate values for the coupling and the scale parameter. Also
the value of $X\sim 175$ $GeV$ should be allowed to be fixed.

The  Fig. \ref{fig5} shows the value of the imaginary part
$V_{im}$ of the potential as a function of $M$ and $X$. Note that
the dependence in $X$ is not rapidly growing, a behavior that if
maintained for large $X$ values and in higher approximations will
indicate an increasing stability of the vacuum being proved, for
the interesting region of high values of $X$.  The picture is for
$f=-(\frac{3}{2})^{2},$ as remarked before, for the positive value
of $f$ the imaginary part of the potential vanish. More generally,
it can be remarked that all the other types of pictures shown in
this section, after being plotted for the positive value of $f$,
show a very similar behavior.
\begin{figure}[ptb]
\begin{center}
\epsfig{figure=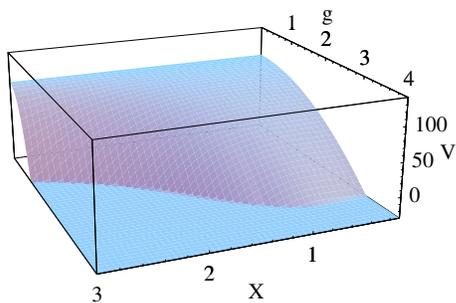,width=6cm}
\end{center}
\caption{   The potential plotted as a function of $X$ and $g$ for
fixed values of $M=333 \ MeV$ and $\mu=6.8\ GeV$. It can be
observed that for each value of $X$, there is critical coupling
$g$ below which the potential becomes negative,
 that is,  lower than its value at vanishing condensates.
  This critical coupling decreases when $X$ grows.}
\label{fig6}
\end{figure}
 Further  Fig.  \ref{fig6} show the dependence of the potential
on the variable $X$  and the gauge coupling $g$. Here the mass
parameter $M$ was fixed to $333$ $MeV$ and again the $\mu $ is
taken as $6.8$ $GeV$ which fixes the minimum in the variable $M$
at $X=0$ to be near the value of $M$. It should be recalled that
we are considering that only one quark is being condensed.
Therefore,  the light quarks which in the present discussion do
not develop their own condensates, should show the observable
value of its constituent masses. Since  this quantity  is fixed by
the value of $M$, the graphics selected to be evaluated are always
chosen to show a minimum near $M=333$ $MeV$ at $X=0$.  The picture
shows how the potential becomes negative (lower than its value at
zero condensate state) when the coupling increases its strength
over an amount  fixed by the value of the quark condensate $X.$
The greater the value of $X$ smaller becomes the critical
coupling.
\begin{figure}[ptb]
\begin{center}
\epsfig{figure=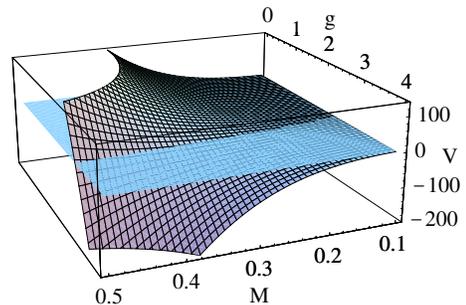,width=6cm}
\vspace{-0.7cm}
\end{center}
\caption{ The  plot of V as a function of $M$ and $g$ for fixed
values of $X=1\ GeV$ and $\mu=6.8\ GeV$.  It shows that there is
critical value of strong coupling $g$ below which the potential at
$M=0$ grow for non vanishing gluon condensates ($M>0$).}
\label{fig7}
\end{figure}
  The Fig.\ref{fig7} show the dependence of the potential on the
gluon condensate and  the coupling constant for fixed values of
$X=1$  and $\mu $= $6.8$ $GeV.$  It can be seen that below certain
critical coupling value near to $2$ for all values of $M,$ the
zero gluon condensate state is stable. Increasing the value of $X$
is not destroying this property and the value of the critical
coupling is simply diminishing for larger $X $ values.

\section{Summary}

The implication of a modified perturbative expansion for QCD are
further investigated. Firstly, an scheme for making well defined
the diagrams of the proposed expansion is introduced. After that,
the zero and one loop contributions to the effective potential are
evaluated. Further, the results of the zero and one loop
calculation are improved by adding the quark condensate dependence
of a relevant two loop term involving the quark propagators.  The
evaluated  potential, in the considered approximation, indicates
an instability of the massless QCD upon the generation of quark
condensates.  At this approximation also there is no terms making
the potential bounded from below. Thus,  next corrections should
produce such terms. Therefore, the results, could be detecting a
possibility for the identification of a sort of Top Condensate
model as possible effective action for massless QCD.  At this
point it seems useful to remark, that the source for the indicated
here effect  could not had been yet detected through numerical
studies, possibly since lattice QCD results are still limited in
the consideration of the fermion determinants. Some questions
concerning the gauge invariance implementation and the
renormalization procedure in the scheme are  commented as well as
some further more detailed investigations expected to be
considered.

\begin{acknowledgments}
The invitation and kind hospitality of the High Energy Section of
the Abdus Salam International Centre for Theoretical Physics
(ASICTP) and its Head Prof. S. Randjbar-Daemi, allowing for two
visits to the Center of one of the authors (AC), is greatly
acknowledged. The Diploma Fellowship at ASICTP of the other author
(DM), also supporting his participation in the work, is also
deeply acknowledged.  The helpful remarks during the development
of the activity of, G. Thompson, V. Skalozub,  M. Fabbrichessi, E.
Pallante, F. Hussain, H. Fritzsch, K. Narain and A. Gonzalez are
also deeply appreciated.
\end{acknowledgments}

\appendix
\section { Regularization  of the singular terms }

The main technical difficulty for the implementation of the
modified  expansion proposed in the works \cite{mpla,
prd,epjc,hoyer, jhep}  is related with the fact that the addition
to the standard free propagator term (let us call it below the
''condensate'' propagator) is  given by a Dirac's Delta function
of the momentum. This circumstance, then can produce singular
diagrams even after the theory is dimensionally regularized. These
singularities are associated to the appearance after some loop
integrals are performed of  Dirac Delta functions,  or standard
Feynman propagators evaluated at zero momentum.  These factors
occurs  due to the conservation of momentum in each vertex .  Let
us consider a vertex with, let us say $n$  ($n=3,4$  for  QCD).
Then, when $n-1$ different condensate lines join to this vertex,
the momentum conservation at it forces the value of the momentum
at the only resting line to vanish.  Therefore, if a condensate
line is attached to this ending, Delta functions evaluated at zero
momentum will appear. On another hand, when a usual free
propagator is connected to this leg the singularity appearing
corresponds to $\frac{1}{p^{2}+i\epsilon }$ evaluated at zero
momentum. This situation should be solved before a full sense
could be given to the modified expansion.  Below,  we propose a
way of considering this problem which clearly should be the
subject of further investigation for its consistency.  Let us
consider the types separately the two types of singularities

\subsection{$\protect\delta (0)$\ $\ $singularities}

A direct idea that can come to the mind after considering the
appearance of these kind of terms is the following.  As we will
employ dimensional regularization, the Delta functions in the
propagators should be also considered as dimensionally regularized
forms of the Dirac Delta function. However, it has been recognized
that similar Delta functions evaluated a zero spacial coordinates
appearing in gravitation theories  can be analytically extended to
continuous dimensions $D$ and moreover, their expression after
taken in the limit of the real space $\ D\rightarrow 4$ \ tends to
vanish.  It can be noticed that this is not a counter-intuitive
result. This is simply because it is possible to impose on the
succession of functions defining the Delta distribution, the
condition of to vanish  at the supporting point without destroying
the possibility that the limit of the integral of any continuous
function multiplied by the elements of the secession, tends to the
value of the function at the support. Therefore, we follow the
same procedure here and interpret the Delta functions appearing as
$D$ dimensional ones. Then, it is possible to step by step
reproduce the arguments of Capper and Leibbrandt \cite{capper} to
conclude  that these factors should vanish after removing the
dimensional regularization. Let us do it below for the sake of
concreteness.

 For the singular $\delta (0)$ we can write
\begin{equation*}
\delta (0)=\int_{E}\frac{dp^{D}}{(2\pi )^{D}}.
\end{equation*}

This is a singular $D$ dimensional integral in Euclidean momentum
space, being completely similar to a one in real space,
considered in \cite {capper}. Then, it can be also written as
follows
\begin{eqnarray*}
\int_{E}\frac{dp^{D}}{(2\pi )^{D}} &=&\int_{E}\frac{dp^{D}}{(2\pi )^{D}}%
\frac{p^{2}}{p^{2}} \\
&=&\int_{0}^{\infty }ds\int_{E}\frac{dp^{D}\ p^{2}}{(2\pi )^{D}}\exp (-s\
p^{2})
\end{eqnarray*}
But, employing the redefinition of the  generalized Gaussian
integral for continuous values of the dimension $D$ constructed in
Ref. \cite{capper}, it is possible to write
\begin{widetext}
\begin{eqnarray*}
\int_{E}\frac{dp^{D}\ }{(2\pi )^{D}}\exp (-s\ p^{2}) &=&\frac{1}{(4\pi )^{%
\frac{D}{2}}}\exp (-s\ f(\frac{D}{2})), \\
\int_{E}\frac{dp^{D}\ p^{2}}{(2\pi )^{D}}\exp (-s\ p^{2}) &=&\frac{1}{(4\pi
)^{\frac{D}{2}}}\left[ \frac{D}{2}s^{-(1+\frac{D}{2})}+s^{-\frac{D}{2}}\ f(%
\frac{D}{2})\right] \exp (-s\ f(\frac{D}{2})),
\end{eqnarray*}
and
\begin{equation*}
\delta (0)=\int_{0}^{\infty }ds\frac{1}{(4\pi )^{\frac{D}{2}}}\left[ \frac{D%
}{2}s^{-(1+\frac{D}{2})}+s^{-\frac{D}{2}}\ f(\frac{D}{2})\right] \exp (-s\ f(%
\frac{D}{2}))
\end{equation*}
\end{widetext}
 where $f$ is the function introduced in \cite{capper}
for extending the generalized Gaussian integral formula for
non-integral dimension arguments. As it should be, these functions
vanish for all integral values of $D.$

Then, after employing the integral definition of the Gamma
function
\begin{equation*}
\Gamma (z)=\int_{0}^{\infty }dt\ t^{z-1}\exp (-t),
\end{equation*}
for the regularized form to the delta function at zero momentum follows
\bigskip
\begin{equation*}
\delta (0)=\frac{f(\frac{D}{2})^{D}}{(4\pi )^{\frac{D}{2}}}\left[ \frac{D}{2}%
\Gamma (-\frac{D}{2})+\Gamma (1-\frac{D}{2})\right] ,
\end{equation*}
which exactly vanish in the limit $D\rightarrow 4^{+}.$

Therefore, we will interpret that the evaluations associated to
the modified expansion are done by using the above representation
for the factors $\delta (0).$ Thus, as a consequence, it will be
considered that  all the diagrams in which only such kind of
singularities appear will vanish in dimensional regularization.

\subsection{b) $\frac{1}{p^{2}+i\protect\epsilon }$ at\ $p=0\ \ $%
singularities}

 For this kind of singular behavior, let us follow the physical
notion about that the modes at zero momentum are appropriately
described only  by the condensate propagator. Therefore, it seems
reasonable to also regularize the dependence of
$\frac{1}{p^{2}+i\epsilon }$\ at exactly zero value of the
momentum in a way that vanish at  this single point in momentum
space $\ p=0.$

Following this idea, let us adopt the following particular
regularization satisfying this criterion
\begin{equation*}
\frac{1}{p^{2}+i\epsilon }{\LARGE \mid }_{reg}=\frac{p^{2}}{%
p^{2}(p^{2}+i\epsilon )+\delta ^{2}}.
\end{equation*}
This expression vanish at $p=0$ and in the limit $\delta
\rightarrow 0$ leads to the scalar Feynman propagator.

 Let us consider now the general loop expansion after assuming
the above two prescriptions and the original free propagators in
the Feynman gauge employed in \cite{jhep}
\begin{align}
G_{g\mu \nu }^{ab}(p,m)& =\frac{\delta ^{ab}g_{\mu \nu }}{p^{2}+i\epsilon }%
-iC\ \delta (p)  \label{free}, \\
G_{q}^{^{f_{1}f_{2}}}(p,M,S)& =-\frac{\delta ^{f_{1}f_{2}}p_{\mu }\gamma
^{\mu }}{p^{2}+i\epsilon \ }+i\text{ }\delta ^{f_{1}f_{2}}C_{f}\ \delta (p),
\notag \\
\chi ^{ab}(p)& =-\frac{\delta ^{ab}}{p^{2}+i\epsilon },  \notag
\end{align}
and the standard vertices of QCD. Then, whenever the $\delta -$%
regularization is employed as described above, it directly follows
that all the diagrams having a fixed number of loops showing both
types of singularities will vanish by taking the limits in both
regulators
\begin{eqnarray*}
\delta  &\rightarrow &0, \\
D &\rightarrow &4.
\end{eqnarray*}

Therefore, after taking the limit $\delta \rightarrow 0,$ before the one $%
D\rightarrow 4,$ the remaining finite diagrams (due to the dimensional
regularization) can be evaluated using the non distorted propagators (\ref
{free}).

At this point is useful to underline some properties of the
original diagrammatic expansion based on the above propagators,
assumed the above explained regularization conditions have
eliminated all the singular contributions. Then it follows:

1) The appearance of a number $m$ of the condensate propagators within a $%
n$-loop diagram will eliminate $m$ of the $n$ loop integrals
associated to this contribution. Therefore, the considered diagram
will be now an ''effective'' $n-m$-loop one.

2) After expressing the condensate parameters $C$ and $C_{f}$ in
favor of the ones:
\begin{align}
m^{2}& =-\frac{6g^{2}C}{(2\pi )^{4}},  \label{newpar} \\
S_{f}& =\frac{g^{2}C_{F}}{4\pi ^{4}}\ C_{f},  \notag
\end{align}
which also incorporate a power of order two of the coupling
constant $g,$ it also follows that the $n$-loop diagrams of the
effective expansion, as
considered as multiple power expansion in the three parameters $m^{2},$ $%
S_{f}$ and $g$,  also shows the property that, given the number of
external legs of the diagram, the number of loops is fixed by the
power of the coupling constant appearing in it. This follows
directly from the fact that each time that a condensate line
appears a loop integral is annihilated and correspondingly the
power of $g$ of the diagram is reduced by two. Let us consider
that $p$ is the power of $g$ corresponding to a $n$ loop diagram
in which the parameters are not redefined. Thus the new power of g
of this diagram in which $m$ condensate lines and the new
parameters are introduced will simply be
\begin{equation}
p^{\prime }=p-2m.  \label{power}
\end{equation}
 This property, then seems to allows for a useful reordering of the
perturbation expansion.  To see elements suggesting it, let us
consider a particular $n$-loop diagram in which the change of the
parameters (\ref{newpar}) have been introduced and corresponding
line symbols have been introduced for the standard and the
condensate propagators separately. Therefore, for any particular
standard type line in this diagram it seems possible to consider
the infinite summation of all the zero order in $g$ (tree)
contributions to the connected propagator (which by construction
have the same number of loops but comes from higher loops in the
original expansion). This is done by considering fixed the other
standard lines. Thus, if not blocked by some difficulty associated
to the combinatorial and symmetry factors in the diagrams,  the
performed infinite additions seems that can be shown to be
possible for all the normal lines. These zero order in the new
expansion propagators are no other things that the expressions
(\ref{quark})-(\ref{quarkcon}) written in the first section. This
was effectively shown in Ref. \cite{jhep}. Thus, it seems possible
to demonstrate that the loop expansion can be reordered to produce
other version of it in which  the new tree propagators will be
(\ref{quark})-(\ref{quarkcon}).  The investigation of this
possibility is expected to be considered in further works.

\section{Appendix B}

Let us comment here the on the question about the implementation
of the gauge invariance in the proposed scheme. This problem can
be divided in two main lines: a) The satisfaction of the
Slavnov-Ward-Takahashi identities given a fixed quantum gauge
condition, and b)  The invariance of the physical quantities under
changes in the quantum gauge condition, in particular for
different values of the gauge parameter $\alpha $ \cite{dewitt}$.$
 Both issues need for additional attention within the modified expansion
under study, since the new appearing elements, the condensate
propagators, being neat distributions, make the discussion more
subtle than in the normal situation.  This study is planned for a
next more basic work. However, some points of interest can be
remarked here below.

  In connection with the satisfaction of the
Slavnov-Ward-Takahashi identities, already in the work
\cite{mpla},  it followed that the one loop correction to the
polarization operator satisfies the simplest identity, that is, is
exactly transverse. This is a non trivial result that suggests the
possibility of its occurrence at higher loops approximations. .
Also in Refs. \cite{hoyer, hoyer1,hoyer2} some general argues and
particular checks of this property in particular  processes were
given. The study of this problem will be continued.

On more general grounds, an observation that also indicates the possibility
of implementing the invariance under the changes of the quantum gauge
condition is the following one.

 It can be noticed that due to the following identities in the sense of the
generalized functions:
\begin{eqnarray*}
p^{2}(\frac{1}{p^{2}+i\epsilon }-iC\text{ }\delta (p)) &=&1, \\
-\gamma _{\mu }p^{\mu }(\frac{-\gamma _{\mu }p^{\mu }}{p^{2}+i\epsilon }%
+iC_{f}\text{ }I\text{ }\delta (p)) &=&I,
\end{eqnarray*}
the propagators of the modified expansion are also inverse kernels
of the second variational derivative of the tree level action of
massless QCD. Therefore,  formal steps can be done to transform
the full generating functional of the Green functions of massless
QCD as written in the Wick expansion representation, to a
functional integral representation over the gluon, quark and ghost
fields. Henceforth, there is the possibility that some of the
changes of variables which are employed to show the quantum gauge
independence of physical quantities could be also implemented in
the case of our interest, since the functional integral will only
differ in the boundary conditions. A particular interesting way of
considering the problem seems to employ  the Yokoyama modification
\cite{yoko} of the Nakanishi-Lautrup $B$ field \cite{naka}
quantization of the interaction free version of massless QCD. In
this scheme the variation of the gauge parameter $\alpha$ can be
implemented as a quantum gauge transformation between the field
operators. Thus, the transformation properties of the free
generating functional associated to the condensate states employed
in Ref. \cite{prd}, under the gauge parameter modifications could
be more efficiently investigated.  These issues requires careful
study that we expect to be able of perform in next works.

\end{document}